\documentclass[aps,prb,amsmath,amssymb,amsfonts,twocolumn,showpacs]{revtex4-1}
\usepackage{ulem}
\usepackage{latexsym}
\usepackage{bm}
\usepackage{graphicx}

\newcommand{\la}{\langle}
\newcommand{\ra}{\rangle}
\newcommand{\be}{\begin{equation}}
\newcommand{\ee}{\end{equation}}
\newcommand{\vk}{\mathbf{k}}

\newcommand{\vq}{\mathbf{q}}
\begin{document}
\title{Theoretical study of X-ray absorption of   three-dimensional topological insulator
$\mathrm{Bi}_2\mathrm{Se}_3$}
\author{Hyun C. Lee}
\email[email:~]{hyunlee@sogang.ac.kr}
\affiliation{Department of Physics and Basic Science Research Institute, Sogang University,
Seoul, Korea}
\date{\today}

\begin{abstract}
X-ray absorption edge singularity which is usually relevant for metals is studied for the 
prototype topological insulator $\mathrm{Bi}_2\mathrm{Se}_3$.
The generalized integral equation  of Nozi\`eres and  Dominicis type for  X-ray edge singularity is derived and  solved.
The spin texture of surfaces states causes a component of singularity dependent on the helicity of the spin texture.
It also yields another component for which the  singularity from excitonic processes is absent.
\end{abstract}
\maketitle
\uline{Introduction}-
The  topological insulators(TI) have been intensively studied recently.\cite{kane-review,zhang-pt}
TI have bulk energy gap but they have conducting (namely gapless) states at \textit{boundary}.\cite{tft}
Quantum Hall state is an example of 2-dimensional TI with broken time reversal symmetry, and the conducting states  
at boundary are nothing but the well-known edge states.\cite{tft}
There exist 3-dimensional TI \textit{with} time reversal invariance, and  they have 
conducting \textit{surface} states (SS) which are protected by $Z_2$ topological invariants in the bulk.\cite{kane3d,balents,roy}
The energy band of the SS takes the form of Dirac cone.\cite{kane-review,tft}
SS have been first observed in $\mathrm{Bi}_{1-x} \mathrm{Sb}_{x}$, but many of their important features were not clearly discerned 
due to  small bulk gap and disorder effect.\cite{hsieh2008}
Stoichiometric TI possessing the simplest SS structure, namely a single Dirac cone, have been proposed for 
$\mathrm{Bi}_2\mathrm{Se}_3$,$\mathrm{Bi}_2\mathrm{Te}_3$,$\mathrm{Sb}_2\mathrm{Te}_3$.\cite{singlecone,model}
The single Dirac cone SS has been observed in ARPES experiment for $\mathrm{Bi}_2\mathrm{Se}_3$.\cite{hasan-single}
These materials can be realized as TI owing to the band inversion mechanism driven 
by large spin-orbit coupling.\cite{singlecone}
The spin texture which is a  distinguished feature of SS  has also been observed in the  spin-resolved ARPES experiment.\cite{hsieh2009}

X-ray absorption (and emission) spectroscopy is a very important method in the study 
of the electronic structure of core electrons.
An incident x-ray photon excites a deep core electron to an unoccupied state with higher energy,
leaving behind a positively charged core hole which can be treated to be immobile in many cases.\cite{ohtaka,mahanbook}
If there exist conduction electrons (of metals), they react to this \textit{suddenly} created  potential by deep core hole.
The conduction electrons interact with deep core hole in two distinctive ways:
the excitonic process\cite{x4}which is essentially attraction between hole and conduction electrons 
and the orthogonality catastrophe\cite{x5} which means a vanishing overlap between the ground-state wavefunctions 
before and after the creation of the deep core hole.
Both excitonic process and orthogonality catastrophe are singular near x-ray absorption edge for fermi liquids,
and they require nonperturbative treatments.\cite{x1,x2,x3,langreth}
For TI, partially filled SS comprise the  conduction electrons in spite of energy gap in the bulk.
Evidently, it is of interest to investigate how the above singular behaviors for the conventional fermi liquids 
are modified for the conducting states realized by SS of TI.

However, there is a caveat.
In the  X-ray absorption experiment with photon incident perpendicular to surface, X-rays  penetrate deep into the bulk of a sample, hence 
providing its bulk properties.\cite{database}
The SS of TI  reside near  surface, so that in this experimental setup the contribution to the absorption
from SS is expected to be rather small.
The decay length (along surface normal) of SS of $\mathrm{Bi}_2 \mathrm{Se}_3$  can be estimated 
to be    in the range of  $4 \sim 10 \,\mathrm{\AA}$  using 
Eq.(32) and Table IV of Ref.[\onlinecite{model}].
For the substantial amount of X-ray absorption to take place in conjunction with SS
the attenuation length of X-ray should be comparable to the decay length  of SS.
The attenuation length of X-ray can be controlled by its energy and the incident angle (measured from surface).
Taking various factors mentioned above into account, 
we choose to focus on a core level  $\mathrm{N}_3 \mathrm{4p}_{3/2}$ of Bi
 whose  binding energy is $678.8\, \mathrm{eV}$.
At this energy the critical angle is 3.26 degrees  and the attenuation length at 3.4 degrees is 
$40 \mathrm{\AA}$ which is indeed
comparable to the decay length of SS.
We note that  the attenuation length for the normal incidence is about 1,000 \AA.\cite{database}
Presently there seems to be no experimental report on the X-ray absorption in  $\mathrm{Bi}_2 \mathrm{Se}_3$.

In this Brief Report we report the results on the X-ray edge problem of $\mathrm{Bi}_2\mathrm{Se}_3$.
The spin texture structure of SS modifies the singular edge  behavior compared to that of conventional Fermi liquids.
The most salient differences are the appearance of a contribution depending on the helicity of spin texture and 
the other which is free of the singularity from excitonic process. 
The main result of this Brief Report is Eq.(\ref{final}).


\uline{Setup}-
The Hamiltonian for the SS is given by (see Eq.(34) of Ref.[\onlinecite{model}], 
and $\alpha, \beta=\uparrow,\downarrow$ denote spin )
\begin{align}
\label{ssH}
\mathsf{H}_{\mathrm{SS}} &= \sum_{\alpha,\beta = \uparrow,\downarrow} c^\dag_{\vk \alpha} \mathsf{h}_{\alpha \beta} c_{\vk \beta},\;
\hat{\mathsf{h}} = ( \mathsf{h}_{\alpha \beta})\cr
\hat{\mathsf{h}} &=(\tilde{C}_0 +\tilde{C}_2 (k_x^2 +k_y^2)) \mathrm{I}_2 +\tilde{A}  (\sigma_x k_y - \sigma_y k_x),
\end{align}
where $\sigma_{x,y}$ are Pauli matrices acting on spin space ($ \mathrm{I}_2$ is a unit matrix) and $c_{\vk \alpha}$ is the destruction operator for SS
with wavenumber $\vk$ and spin $\alpha$.
In Eq.(\ref{ssH}) we have ignored the trigonal distortion terms proportional to 
$( k_x  \pm i k_y)^3$.
The numerical values of the parameters of Eq.(\ref{ssH}) which are appropriate for $\mathrm{Bi}_2 \mathrm{Se}_3$ are given by
$\tilde{C}_0 =3.37 \times 10^{-2} \,\mathrm{eV}$,
$\tilde{C}_2 = 23.7 \, \mathrm{eV} \cdot \text{\AA}^2$, 
$\vert \tilde{A} \vert = 3.30 \, \mathrm{eV} \cdot \text{\AA}$.
The Fig.4 of Ref.[\onlinecite{model}] suggests the wavenumber cutoff be 
$k_{\mathrm{c}} \approx 0.1  \text{\AA}^{-1}$.
The energy eigenvalue is given by
\be
E_\pm(\vk) = \tilde{C}_0 + \tilde{C}_2 k_\parallel^2 \pm \vert \tilde{A} \vert k_\parallel,\;k_\parallel = \sqrt{k_x^2 + k_y^2}.
\ee
When the $\tilde{C}_2 k_\parallel^2$ is much smaller than
$ \vert \tilde{A} \vert k_\parallel$ the Dirac cone structure with apex at $\vk =0$ is manifest.

The non-interacting Matsubara Green's function of SS can be expressed as (the hat  denotes matrix
and $\phi$ is the azimuth angle in x-y plane)
\be
\label{green0}
 \hat{g}(i\epsilon,\vk)=  \mathrm{I}_2 \, g_{\mathrm{d}}
+  \Big [(+i) \hat{e}_{12} e^{-i \phi} + (-i) \hat{e}_{21} e^{i \phi} \Big ]\, g_{\mathrm{o}},
\ee
where $\hat{e}_{ij}$ is a 2x2 matrix whose only non-vanishing element is 1  at $(i,j)$ entry, and 
\begin{align}
\label{green1}
 g_{\mathrm{d}}(i\epsilon,\vk) &= \frac{ i\epsilon + \mu - \tilde{C}_0 -\tilde{C}_2 k^2_\parallel}{
[i \epsilon+\mu - E_+(\vk)][i \epsilon+\mu - E_-(\vk)]}, \cr
 g_{\mathrm{o}}(i\epsilon,\vk) &= \frac{ \tilde{A} k_\parallel}{
[i \epsilon+\mu - E_+(\vk)][i \epsilon+\mu - E_-(\vk)]},
\end{align}
where $\mu$ is the chemical potential,  and $\mu > \tilde{C}_0$ will be assumed (namely, the
lower Dirac cone is completely occupied). The interrelation between the spin and the angle in
 Eq.(\ref{green0}) is nothing but the manifestation of the spin texture of SS.\cite{model}
Note that the Green's functions of Eq.(\ref{green1}) are independent of the angle $\phi$.
The Green's functions summed over wavenumber in the long time limit are given by ($\tau$ is imaginary time)
\be
\label{Sign}
g_{\mathrm{d}}(\tau) =-\frac{\rho}{\tau},\;\;
 g_{\mathrm{o}}(\tau)= \mathrm{sgn}(\tilde{A}) \, g_{\mathrm{d}}(\tau),
\ee
where $\rho = A k_F / 4 \pi v_F$ is the density of states at Fermi energy             
 ($A$ is the area of unit cell),
and $v_F$ and $k_F$ is the Fermi velocity and Fermi momentum, respectively, 
whose detailed form does not concern us here.
Note the sign factor $\mathrm{sgn}(\tilde{A})$ in Eq.(\ref{Sign}), which is the signature of the \textit{helicity} of the 
spin texture of SS.\cite{model}

The core level  $\mathrm{N}_3 \mathrm{4p}_{3/2}$ of Bi is  labeled by the 
z-component of the total angular momentum $J=3/2$.
\be
\label{hole}
\mathsf{H}_{\mathrm{hole}} = E_h \sum_{m_J= \pm 3/2,\pm 1/2} b^\dag_{m_J} b_{m_J},
\ee
where $b^\dag_{m_J}$ is the creation operator of the hole. $E_h$ is the core level energy, and $\mu+E_h$ is
the (unrenormalized) threshold energy for X-ray absorption.
The potential created by the deep core hole will be assumed to be spherically symmetric, and 
for simplicity we will consider the isotropic scattering only, so that the potential scattering 
matrix element for SS is simplified to\cite{x3}
\be
\label{s-wave}
V_{\vk \vk'} = - V_0, \quad V_0 > 0  \;\; \text{is constant}.
\ee
Then the interaction Hamiltonian between SS and the deep core hole is given by
\be
\label{interaction}
\mathsf{H}_{\mathrm{int}} = \sum_{\vk,\vk'} (-V_0) 
(\sum_{\alpha} c^\dag_{\vk \alpha} c_{\vk' \alpha} ) (\sum_{m_J} b^\dag_{m_J} b_{m_J}),
\ee
where a suitable cutoff in the wavenumber sum is assumed implicitly.
The total Hamiltonian consists of 
\be
\label{total}
\mathsf{H}_{\mathrm{tot}} = \mathsf{H}_{\mathrm{SS}}+\mathsf{H}_{\mathrm{hole}}+\mathsf{H}_{\mathrm{int}}.
\ee
With the Hamiltonian Eq.(\ref{total}) the hole quantum number $m_J$ is conserved, 
and it implies that the deep core hole Green's function is given by
\be
\label{d-function}
-\langle b_{m_J}(\tau) b^\dag_{m_J^\prime}(\tau') \rangle 
= \delta_{m_J,m_J^\prime} D(\tau-\tau'),
\ee
where the function $D(\tau)$ is independent of $m_J$.
The X-ray absorption intensity $I(\omega)$ ($\omega$ is the frequency of the incident X-ray) 
can be expressed in terms of  correlation function [using Eq.(\ref{d-function})] as follows:\cite{doniach}
\begin{align}
\label{correlation}
&I(\omega )    = \mathrm{Im}\, \int_0^\infty e^{i \omega \tau} F(\tau) \Big \vert_{i \omega \to \omega + i \delta}, \cr
&F(\tau) = \sum_{\vk,\vk',\alpha,\beta} \sum_{m_J} M_{\vq \lambda}(\vk,\alpha \vert m_J)
 M^*_{\vq \lambda}(\vk',\beta \vert m_J) \cr
&\times F_{\vk \vk' \alpha \beta \vert m_J}(\tau), \cr
&F_{\vk \vk' \alpha \beta \vert m_J}(\tau) =
\la  c_{\vk\alpha}(\tau) b_{m_J}(\tau)  b_{m_J}^\dag(0)c^\dag_{\vk'\beta}(0)  \ra,
\end{align}
where $M_{\vq \lambda}(\vk,\alpha \vert m_J)$ is the X-ray transition matrix element from the deep core state $\vert m_J \ra$ 
to the SS (Bloch state) $\vert \vk,\alpha \ra$, and $\vq$ and $\lambda$ is the wavenumber and the polarization of the 
incident X-ray, respectively.
In many cases of interest, the wavenumber ($\vq,\vk$) dependence  of the X-ray transition matrix element 
can be ignored. 
This is due to the localized nature of the wave function of core electron.
In the presence of the strong spin-orbit coupling such as the case of  Bi,
the electron-photon interaction receives additional
contribution from the spin-orbit coupling.\cite{blume}
The explicit form of the transition matrix element including spin-orbit contribution is 
\begin{align}
\label{transition}
& M_{\vq \lambda}(\vk \alpha \vert  m_J ) =
\int d^3 \vec{r} \, e^{i \vq \cdot \vec{r}} \,\Big \{
\frac{(-e)}{m_e}  \phi_{\vk}^\dag \big[-i \hbar \vec{e}_{\vq \lambda} \cdot\nabla \Psi_{m_J}\big ] \cr
&+\frac{\hbar (-e)}{4 m_e^2 c^2}  \phi_\vk^\dag \vec{\sigma}\cdot
\big[- \hbar \omega \vec{e}_{\vq \lambda} \times \nabla \Psi_{m_j}
+ \nabla V \times \vec{e}_{\vq \lambda}  \Psi_{m_J}\big]\Big \},
\end{align}
where $\vec{e}_{\vq \lambda}$ is the polarization vector of X-ray and $V(\vec{r})$ is the periodic crystal potential.
$\phi_\vk(\vec{r})$ and $\Psi_{m_J}(\vec{r} ) $  is the (spinor) Bloch wavefunction of SS and 
 the (spinor) wavefunction of core electron, respectively.
The dipole approximation   $ e^{i \vq \cdot \vec{r}} \approx 1$ will be assumed below.

\uline{Correlation functions}-
The correlation function $F_{\vk \vk' \alpha \beta \vert m_J}(\tau)$ of Eq.(\ref{correlation}) can be obtained from 
\begin{align}
\label{F-function}
F_{\vk \vk' \alpha \beta \vert m_J}(\xi,\xi' \vert \tau_1,\tau_2) = 
\la  T c_{\vk\alpha}(\xi) b_{m_J}(\tau_1)  b_{m_J}^\dag(\tau_2)c^\dag_{\vk'\beta}(\xi')  \ra
\end{align}
by the limiting procedure
 $ \xi \to \tau_1 -\tau_c$ and $\xi' \to \tau_2 + \tau_c$ ($\tau_c$ is a short-time cutoff).
$T$ denotes time ordering.
For the absorption we have to take $\tau_1 > \tau_2$.
We employ the equation of motion method of Ref.[\onlinecite{langreth}] to derive the integral equation for 
$F_{\vk \vk' \alpha \beta \vert m_J}(\xi,\xi' \vert \tau_1,\tau_2)$.
The conservation of the  hole number causes the equation of motion to close on itself,\cite{langreth} and we find 
the integral equation
\begin{align}
\label{equation1}
& F_{\vk \vk' \alpha \beta \vert m_J}(\xi,\xi' \vert \tau_1,\tau_2)   = \delta_{\vk \vk'}
g_{\alpha \beta}(\xi-\xi',\vk)  D(\tau_1-\tau_2) \cr
&+\sum_{\vq',\gamma} \int^{\tau_1}_{\tau_2} d \tau g_{\alpha \gamma }(\xi-\tau,\vk) V_{\vk \vq'} 
F_{\vq' \vk'\gamma \beta \vert m_J}(\tau,\xi' \vert \tau_1,\tau_2),
\end{align}
where $g_{\alpha \beta}(\xi,\vk)$ is the Green's function  Eq.(\ref{green0}) in (imaginary) time domain.
Decomposing  $F_{\vk \vk' \alpha \beta \vert m_J}$ as follows
\be
\label{product}
  F_{\vk \vk' \alpha \beta \vert m_J}(\xi,\xi' \vert \tau_1,\tau_2)= 
G_{\vk \vk' \alpha \beta}(\xi,\xi' \vert \tau_1,\tau_2)   D(\tau_1-\tau_2)
\ee
Eq.(\ref{equation1}) becomes the following integral equation:
\begin{align}
\label{equation2}
& G_{\vk \vk' \alpha \beta}(\xi,\xi' \vert \tau_1,\tau_2) = \delta_{\vk \vk'}\,g_{\alpha \beta}(\xi-\xi',\vk)  \cr
&+\sum_{\vq',\gamma}\int^{\tau_1}_{\tau_2} d \tau g_{\alpha \gamma }(\xi-\tau,\vk) V_{\vk\vq'} 
G_{\vq' \vk'\gamma \beta}(\tau,\xi' \vert \tau_1,\tau_2).
\end{align}
Eq.(\ref{equation2}) is the generalization of Eq.(17a) in Ref.[\onlinecite{x3}] to our case of SS.
$G_{\vk \vk'\alpha \beta}$ and $D(\tau_1-\tau_2)$ of Eq.(\ref{product}) represents 
the excitonic processes and the orthogonality catastrophe, respectively.\cite{x3}
It can be shown that the hole Green's function $ D(\tau_1-\tau_2)$ can be obtained from the solution of 
Eq.(\ref{equation2}) via parametric integral (see Eq.(21) of Ref.[\onlinecite{x3}] and Eq.(11) of Ref.[\onlinecite{langreth}]).
Thus once Eq.(\ref{equation2}) is solved, we can find X-ray absorption intensity from Eqs.(\ref{correlation},\ref{product}).

In fact, we need to find the Green's function $G_{\vk \vk' \alpha \beta}$ summed over wavenumber weighted by 
transition matrix element $M_{\vq \lambda}(\vk \alpha \vert m_J)$. In most cases of simple metals the wavenumber dependence 
of transition matrix element is ignored. 
However, in our case such dependence is crucial because, as can be seen in Eq.(\ref{green0}),
the spin texture structure is encoded in the angle dependence of Green's function.
In view of this we expand the transition matrix element in Fourier series of $e^{i \phi}$ but we will ignore
$k_\parallel=\sqrt{k_x^2+k_y^2}$ dependence.
\be
M_{\vq \lambda}(\vk \alpha \vert m_J) \approx \sum_{n=0,\pm 1} e^{i n\phi} 
M_{\vq \lambda}^{(n)}( \alpha \vert m_J),
\ee
where only $n=0,\pm 1$ terms are kept since  the higher order contributions 
will be smaller because they  involve higher power of  $ k_\parallel r $.
Let us define (henceforth time arguments are suppressed for notational clarity)
\be
\label{average1}
\bar{G}_{\alpha\beta \vert n, n'} \equiv  \sum_{\vk,\vk'} e^{i n \phi}
(e^{i n \phi'})^* G_{\vk \vk'  \alpha\beta},
\ee
where a cutoff in wavenumber sum is implicitly assumed.
Applying the definition Eq.(\ref{average1}) to Eq.(\ref{green0}) we find
\begin{align}
\label{zero}
 \bar{g}_{\alpha \beta \vert 0, 0} &= \bar{g}_{\alpha \beta \vert 1, 1}=\bar{g}_{\alpha \beta \vert -1, -1}=
\delta_{\alpha \beta} g_{\mathrm{d}}(\tau), \cr
\bar{g}_{\alpha \beta \vert 1, 0} &=
\bar{g}_{\alpha \beta \vert 0, -1}= \delta_{\alpha \uparrow} \delta_{\beta \downarrow}  (+i)   g_{\mathrm{o}}(\tau), \cr
\bar{g}_{\alpha \beta \vert 0,1} &=
\bar{g}_{\alpha \beta \vert -1, 0}= \delta_{\alpha \downarrow} \delta_{\beta \uparrow}  (-i)   g_{\mathrm{o}}(\tau),\cr
\bar{g}_{\alpha \beta \vert 1, -1}&= \bar{g}_{\alpha \beta \vert -1, 1}=0.
\end{align}
Now Eq.(\ref{equation2}) can be recast into the following form [ recall Eq.(\ref{s-wave}) ]
\begin{align}
\label{equation3}
& \bar{G}_{\alpha\beta \vert n, n'}(\xi,\xi' \vert \tau_1,\tau_2)   =  \bar{g}_{\alpha \beta \vert n, n'}(\xi-\xi') \cr
&+(-V_0) \int_{\tau_2}^{\tau_1} d \tau \bar{g}_{\alpha \gamma \vert n, 0} (\xi -\tau) 
\bar{G}_{\gamma \beta \vert 0, n'} (\tau,\xi' \vert \tau_1,\tau_2),
\end{align}
which are  coupled integral equations. Noting the factor $ \bar{g}_{\alpha \gamma \vert n, 0} $ and from
Eq.(\ref{zero}), we find the nontrivial   solutions (with non-vanishing 2nd term) obtain only for  $n = 0, \pm 1$.

\uline{Solution of integral equation}-
The well-known  Nozi\`eres and Dominicis (ND) (asymptotic) solution 
in long time limit is [$ g(\xi) = -\frac{\rho}{\xi}$=non-interacting Green's function] \cite{x3}
\be
G_{\mathrm{ND}}(\xi,\xi' \vert \tau_1,\tau_2) = \cos^2 \delta g(\xi-\xi') \left[
\frac{(\xi-\tau_2)(\tau_1 - \xi')}{(\tau_1-\xi)(\xi'-\tau_2)}\right]^{\delta/\pi}
\ee
which satisfies [ compare with Eq.(\ref{equation3}) ]
\begin{align}
\label{NDequation}
& G_{\mathrm{ND}}(\xi,\xi' \vert \tau_1,\tau_2)   = 
 g(\xi-\xi') \cr
&+(-V_0) \int_{\tau_2}^{\tau_1} d \tau g (\xi -\tau) 
G_{\mathrm{ND}} (\tau,\xi' \vert \tau_1,\tau_2),
\end{align}
where $\delta$ is the s-wave scattering phase shift
\be
\delta = \tan^{-1}[\pi V_0 \rho].
\ee

For $n=n'=0$, using Eq.(\ref{zero}),  Eq.(\ref{equation3}) is found to  reduce to Eq.(\ref{NDequation}).
Hence (time arguments suppressed)
\be
\label{sol1}
\bar{G}_{\alpha\beta \vert 0,0} = \delta_{\alpha \beta} G_{\mathrm{ND}}.
\ee
Noting Eqs.(\ref{Sign},\ref{zero}) we also find the solutions for $n'= \pm 1$
\begin{align}
\label{zerozero}
\bar{G}_{\alpha \beta \vert 0, 1} &= \delta_{\alpha \downarrow} \delta_{\beta \uparrow} (-i) \mathrm{sgn}(\tilde{A}) G_{\mathrm{ND}}, \cr
\bar{G}_{\alpha \beta \vert 0, -1} &= \delta_{\alpha \uparrow} \delta_{\beta \downarrow} (+i) \mathrm{sgn}(\tilde{A}) G_{\mathrm{ND}}.
\end{align}
For other values of $n'$, $\bar{G}_{\alpha \beta \vert 0 n'}=0$.

Next consider the case of $n=1$. From the property of $g_{\alpha \gamma \vert 1,0}$ [see Eq.(\ref{zero})]
the nontrivial solutions obtain only for $\alpha =\uparrow$, so that
\be
 \bar{G}_{\downarrow \beta \vert 1, n'} = \bar{g}_{\downarrow \beta \vert 1 ,n'}.
\ee
Now take $\alpha = \uparrow$.
\begin{align}
\label{equation5}
& \bar{G}_{\uparrow \beta \vert 1, n'}(\xi,\xi' \vert \tau_1,\tau_2)   =  \bar{g}_{\uparrow \beta \vert 1, n'}(\xi-\xi') \cr
&+(-V_0) \int_{\tau_2}^{\tau_1} d \tau \bar{g}_{\uparrow \downarrow \vert 1, 0} (\xi -\tau) 
\bar{G}_{\downarrow \beta \vert 0, n'} (\tau,\xi' \vert \tau_1,\tau_2).
\end{align}
If $n'= -1$, then from $ \bar{g}_{\uparrow \beta \vert 1, -1} =0$ and 
$\bar{G}_{\downarrow \beta \vert 0, -1}=0$ [see Eq.(\ref{zerozero})],  we conclude that 
\be
\bar{G}_{\uparrow \beta \vert 1,-1} =0.
\ee
For the case of  $n' =0$ of Eq.(\ref{equation5}),
$\beta$ should be $\downarrow$, otherwise both $\bar{g}$ and $\bar{G}$ vanish. Thus
\be
\bar{G}_{\uparrow \uparrow \vert 1 0} =0
\ee
and for $\beta = \downarrow$, multiplying both sides of  Eq.(\ref{equation5}) by $(-i) \mathrm{sgn}(\tilde{A})$,
we find the equation becomes exactly ND type Eq.(\ref{NDequation}), so that
\be
\bar{G}_{\uparrow \downarrow \vert 1 0} = (+i) \mathrm{sgn}(\tilde{A}) G_{\mathrm{ND}}.
\ee
Repeating the similar analysis for other cases we obtain
\begin{align}
\label{result1}
&\bar{G}_{\alpha \beta \vert 0, 0} = \delta_{\alpha \beta} G_{\mathrm{ND}}, \;\;
\bar{G}_{\alpha \beta \vert 1,  -1} = \bar{G}_{\alpha \beta \vert -1,1}=0, \cr
&\bar{G}_{\alpha \beta \vert 11} =\left( \begin{matrix}  G_{\mathrm{ND}}   & 0 \cr
                                                   0 & g_{\mathrm{d}}  \end{matrix}
                                   \right), \;\;
\bar{G}_{\alpha \beta \vert -1 -1} =\left( \begin{matrix}  g_{\mathrm{d}}   & 0 \cr
                                                   0 &   G_{\mathrm{ND}} \end{matrix}
                                   \right), \cr
&\bar{G}_{\alpha \beta \vert 01}=\bar{G}_{\alpha \beta \vert -1 0} = \left ( \begin{matrix} 0 & 0 \cr
                                                                             -i \mathrm{sgn}(\tilde{A}) G_{\mathrm{ND}} & 0 
                                                                            \end{matrix} \right), \cr
&\bar{G}_{\alpha \beta \vert 10}=\bar{G}_{\alpha \beta \vert 0 -1 } = \left ( \begin{matrix} 0 & i \mathrm{sgn}(\tilde{A}) G_{\mathrm{ND}} \cr
                                                                             0 & 0 
                                                                            \end{matrix} \right).
\end{align}
The leading behavior of the core hole Green's function $D(\tau)$ 
can be obtained from the second order linked-cluster expansion.\cite{mahanbook}
The important contribution turns out to be
\be
\label{hole-leading}
V_0^2 \, \int_0^\tau d \xi  \int_0^\tau  d \xi' \mathrm{Tr} \Big[ g^{(0)} (\xi-\xi')   g^{(0)} (\xi' -\xi) \Big ],
\ee
where $ g^{(0)} (\xi-\xi') = \sum_\vk g(\vk,\xi-\xi') = g_{\mathrm{d} }(\xi-\xi') \mathrm{I}_2$.
Thus the helicity of the spin texture of SS does not affect the core hole Green's function, so that the situation 
becomes essentially identical with that of ND solution.
The evaluations of the integral of Eq.(\ref{hole-leading}) yields logarithms, which is to be exponentiated in the linked-cluster
expansion. Comparing with ND solution we find ($N_c =2$, ($\uparrow,\downarrow$))
\be
\label{result2}
D(\tau > 0) \sim e^{-\omega_T^* \tau} \frac{1}{(\tau/\tau_c)^{N_c (\delta/\pi)^2}}.
\ee
$\omega_T^*$ is the renormalized threshold for X-ray absorption.

\uline{Results}-Combining the solutions Eqs.(\ref{result1},\ref{result2}) 
with the transition matrix element Eq.(\ref{transition})
we can obtain the result for X-ray absorption intensity.
 Eq.(\ref{transition}) has two components: one from direct dipole transition and the other from spin-orbit coupling.
Dipole transition conserves the spin, so that it is diagonal in spin (namely, 
$M_\alpha M^*_\beta \propto \delta_{\alpha \beta}$. This can be verified explicitly in our case).
Then Eq.(\ref{result1}) tells us that only $\bar{G}_{\alpha \beta \vert (00),(11),(-1,-1)}$ contribute. 
Among these,
$G_{\mathrm{ND}}$ includes the singularity from excitonic processes while $g_d$ does not. The cross term of dipole transition
and the spin-orbit contribution allows spin off-diagonal configuration. From Eq.(\ref{result1}) we find that 
this contribution is proportional to the helicity $\mathrm{sgn}(\tilde{A})$.
The above considerations give ($\Theta(x)$ is step function)
\begin{align}
\label{final}
I(\omega) &\sim \Theta(\omega - \omega_T^*) \Big[   c_{\mathrm{d}}^\prime (\omega - \omega_T^*)^{N_c (\delta/\pi)^2} \cr
&+(c_{\mathrm{d}} +  \mathrm{sgn}(\tilde{A}) c_{\mathrm{o}} )
(\omega - \omega_T^*)^{-2 \delta/\pi + N_c (\delta/\pi)^2} \Big],
\end{align}
where $ c_{\mathrm{d}}^\prime,c_{\mathrm{d}}, c_{\mathrm{o}}$ are constants.

\uline{Summary and concluding remarks}-We have studied the X-ray absorption edge singularity of the prototype TI,
$\mathrm{Bi}_2\mathrm{Se}_3$. For the singularity to exist, gapless conducting states are necessary, and  SS
provide those.  Due to the spin texture of SS, two interesting modifications compared to that of
convention metals arise:(1) helicity dependent contribution  [$\mathrm{sgn}(\tilde{A})$ term of Eq.(\ref{final})]
(2) contribution free of singularity from excitonic process [
the first term of Eq.(\ref{final})].
These features can be verified experimentally by extracting the surface contributions 
using the angle and energy dependence of the penetration depth of incident X-ray, 
and we have also suggested a specific core level appropriate for experiments.
The angle and energy dependence of the penetration depth can be also used in distinguishing the helicity effect of TI 
from that  of the conventional Rashba spin-orbit energy bands since the latter reside in the bulk.
We mention in comparison that the graphenes have  \textit{two} Dirac cones from valley structure, so that its 
qualitative properties different from those of TI.\cite{mine1} We also mention that when the Fermi energy
crosses the apex of Dirac cone, all of the singularities disappear.\cite{mine1}  
However, this situation is not generic for TI and is not elaborated in this Report.

\begin{acknowledgments}
This research was supported by Basic Science Research Program
through the National Research Foundation of Korea(NRF) funded
by the Ministry of Education, Science and Technology(No.2010-0025406) and 
by  Mid-career Researcher Program through NRF grant funded by 
the Mest (No. 2010-0000179) and by the Special Research Grant of
Sogang University.
We are grateful to H. Kim for useful comments.
\end{acknowledgments}

\end{document}